\documentclass[conference]{IEEEtran}
\IEEEoverridecommandlockouts
\usepackage[T1]{fontenc}
\usepackage{cite}
\usepackage{amsmath,amssymb,amsfonts}
\usepackage{algorithmic}
\usepackage{graphicx}
\usepackage{textcomp}
\usepackage{xcolor}
\def\BibTeX{{\rm B\kern-.05em{\sc i\kern-.025em b}\kern-.08em
    T\kern-.1667em\lower.7ex\hbox{E}\kern-.125emX}}
\begin{document}

\title{SENSOR-BASED NATURAL FREQUENCY TESTING\\
%{\footnotesize \textsuperscript{*}Note: Sub-titles are not captured in Xplore and should not be used}
%\thanks{}
}

\author{\IEEEauthorblockN{1\textsuperscript{st} Nicholas Suits}
\IEEEauthorblockA{\textit{Electrical and Computer Eng, Clarkson University} \\
\textit{8 Clarkson Ave, Potsdam, NY.}\\
suitsn@clarkson.edu}
\and
\IEEEauthorblockN{2\textsuperscript{nd} Masudul Imtiaz}
\IEEEauthorblockA{\textit{Electrical and Computer Eng, Clarkson University} \\
\textit{8 Clarkson Ave, Potsdam, NY.}\\
mimtiaz@clarkson.edu}
}

\maketitle%\documentclass[journal,article,submit,pdftex,moreauthors]{Definitions/mdpi} 
%\firstpage{1} 
%\makeatletter 
%\setcounter{page}{\@firstpage} 
%\makeatother
%\pubvolume{1}
%\issuenum{1}
%\articlenumber{0}
%\pubyear{2026}
%\copyrightyear{2026}
%\usepackage{graphicx}
%\externaleditor{Firstname Lastname} % More than 1 editor, please add `` and '' before the last editor name
%\datereceived{ } 
%\daterevised{ } % Comment out if no revised date
%\dateaccepted{ } 
%\datepublished{ } 
%\Title{SENSOR-BASED NATURAL FREQUENCY TESTING}
% Author Orchid ID: enter ID or remove command
%\newcommand{\orcidauthorA}{0000-0000-0000-000X} % Add \orcidA{} behind the author's name
%\newcommand{\orcidauthorB}{0000-0000-0000-000X} % Add \orcidB{} behind the author's name
%\Author{Nicholas Suits and Masudul Imtiaz}
% MDPI internal command: Authors, for metadata in PDF
%\AuthorNames{Nicholas Suits and Masudul Imtiaz}
%\longauthorlist{yes}
% Affiliations / Addresses (Add [1] after \address if there is only one affiliation.)
%\address{%
 %\quad Clarkson University, Potsdam, NY \\suitsn@clarkson.edu and mimtiaz@clarkson.edu}

% Current address and/or shared authorship
%\firstnote{Current address: Affiliation.}  
% Current address should not be the same as any items in the Affiliation section.

%\secondnote{These authors contributed equally to this work.}
% The commands \thirdnote{} till \eighthnote{} are available for further notes.

%\simplesumm{} % Simple summary

%\conference{} % An extended version of a conference paper

% Abstract (Do not insert blank lines, i.e. \\) 

\begin{abstract}
Everything that exists has a natural frequency; this material characteristic is something that must be known and fully understood. If we fail to predict, measure, and address potential natural frequency concerns, it could significantly reduce the life span of our equipment or cause it to fail immediately when put into service. There are a few methodologies used to study natural frequencies, one being computer simulations and the other being physical tests done on the equipment. In this paper, we will focus on testing natural frequencies and discuss how we measure our excitation, our form of excitation, the type of data we are able to export, as well as what we are able to do with that data. These principles can be applied to any type of machinery or object where vibration could be of concern. For our purposes, we will primarily focus on rotating machinery, such as generators, gearboxes, and motors.
\end{abstract}


%\abstract {Everything that exists has a natural frequency; this material characteristic is something that must be known and fully understood. If we fail to predict, measure, and address potential natural frequency concerns, it could significantly reduce the life span of our equipment or cause it to fail immediately when put into service. There are a few methodologies used to study natural frequencies, one being computer simulations and the other being physical tests done on the equipment. In this paper, we will focus on testing natural frequencies and discuss how we measure our excitation, our form of excitation, the type of data we are able to export, as well as what we are able to do with that data. These principles can be applied to any type of machinery or object where vibration could be of concern. For our purposes, we will primarily focus on rotating machinery, such as generators, gearboxes, and motors.}

% Keywords
%\keyword{Accelerometer; Impulse Hammer; Natural Frequency Testing; Ping Testing; Shaker Testing; Vibration Analysis; } 

%%%%%%%%%%%%%%%%%%%%%%%%%%%%%%%%%%%%%%%%%%

% The order of the section titles is different for some journals. Please refer to the "Instructions for Authors” on the journal homepage.

\section{Introduction}

Nearly everything in the engineering world can be modeled or predicted using various types of software and calculations. However, these models can only get you so far due to the real-world conditions that exist \cite{mahn1999impact}. Although these models also assume that the part is created to the exact specification laid out in the drawing, and there is no degradation of materials over time. Through testing, we are able to compare the predicted measurements with the actual field conditions. Natural frequency testing gives us an idea of how an object will react under certain running conditions \cite{ewins2010control}. From here, as engineers, we are able to look at problematic frequencies and modify our equipment. To optimize the equipment to have a longer time in service, or find problem areas that would have developed over time \cite{guo2023effect}. Something that a model would not have the capability of telling you. 

Owners of rotating machinery, in particular, have increased interest in this vibration analysis. In a rotating machine, the modes get excited when your rotor spins at certain resonant frequencies. Your running speed, whether fixed or variable, could affect the vibration levels of your machine \cite{ewins2010control}. The modes on your way to that running speed, although less impactful, are also important to understand. Especially if you plan to do predefined holds on your way to your running speed. Although some rotating machines may not have vibration trip limits, or the vibration of the machine does not trip the limits. Extended periods of time that are exciting for these modes and vibrating the machine will reduce the longevity of the equipment or cause an increase in unscheduled downtime \cite{salawu1997detection}.   

Testing before starting up is a way that manufacturers can validate the construction and make modifications to their unit before their first start-up. Objects do have a natural frequency; however, they can be modified through the correct engineering actions, making this step very important, especially if models show a natural frequency that is close to the intended operating speed \cite{ccakar2018method}. The added benefit is to have data that you are able to reference down the road, as the pre-start-up condition should be the optimal condition of the machine. Once significant run time is seen on the unit, we are then able to go back and re-test to compare the results to see any discrepancies. Even without these initial start-up data, testing a machine that has been in service for several years could be very telling of its condition. Overtime components in rotating machines wear down and could degrade. Making testing of these machines imperative to determine equipment health and to determine if any corrective measures are needed \cite{hassiotis1993assessment}.

Natural frequency testing can also be just as important to determine machine downs, even if you do not have the pre-startup data. With an Operating Deflection Shape (ODS) model, we are able to see where a piece of equipment may have a problem area \cite{kessler2002damage}. In an ODS model utilizing different accelerometer points around the equipment, we can create a 3-dimensional shape that moves as the equipment would at any given frequency. Doing so gives the analyzer the ability to view if certain components are in or out of phase with one another, showing an issue with the equipment \cite{kessler2002damage}. This type of analysis is used on both new and old types of equipment to determine if the part was made correctly or if it was damaged over the years of use.       
	
In this paper, we will examine different aspects of natural frequency testing and how these variables could impact our data. To do this, we will be examining a multi-stage gearbox that is comprised of two separate gearbox pedestals. This gearbox sits on a foundation with three separate interfaces. The concrete foundation, grout plate, steel base plate, and then the gearbox itself; for each of these interfaces, a mechanical feature binds them together. We will be examining the interface connections from the foundation to see if these multiple interface connections could cause any variation in the natural frequencies we measure. We will also be looking at the significance of the excitation location on your test piece. Take a look at the comparison of vibration data taken with an accelerometer vs a velometer. While finally looking at the impact of noise levels during the test, by comparing data with local pumps on vs off. Aside from the data, we will also look at other ways to run natural frequency testing and look at the feasibility and criteria of how to choose the correct equipment.

%%%%%%%%%%%%%%%%%%%%%%%%%%%%%%%%%%%%%%%%%%
\section{Measurement Devices}

The main type of instrument used in natural frequency testing is a piezoelectric accelerometer. These devices use motion as a means to convert mechanical movement into electrical signals \cite{wu2023research}. To do this, the accelerometer has a small mass inside the casing of the instrument. When the device is excited, the mass then begins to vibrate, which puts a force on the piezoelectric element. This element then produces a charge, which gets fed into a built-in amplifier \cite{wu2023research}. This amplifier then sends out a voltage signal, which can be measured by the data acquisition system. The user is then able to translate the voltage signal vibration data \cite{wu2023research}.  

This type of accelerometer can be used in a single-axis configuration where only a single plane is measured. Or in a tri-axial configuration where the component is measured in three directions \cite{wu2023research}. A tri-axial accelerometer has advantages as you have one device that is able to detect excitation in all three planes of the test piece. Whereas a single-axis accelerometer is used when there is a single plane in question \cite{wu2023research}. For this reason, single-axis accelerometers are typically used in equipment monitoring applications, whereas triaxial accelerometers are used in natural frequency testing. In the data gathered on the gearbox, we used triaxial accelerometers \cite{wu2023research}. Specifically, we used a PCB Piezotronics model 356A16 triaxial accelerometer \cite{PCBACC}. This allowed us to excite and record data in all axes of the gearbox to understand its movement and natural frequency response. 

\begin{figure}[htbp]
\centering
\includegraphics[width=0.4 \linewidth]{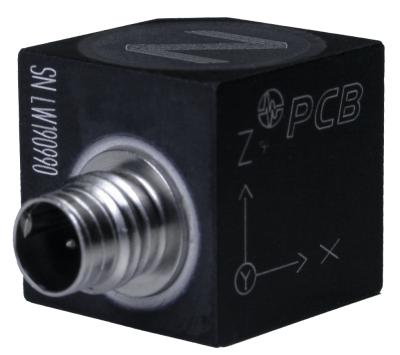}
\caption{Piezotronics model 356A16 triaxial accelerometer, this is the same model that was utilized in our gear box testing. \cite{PCBACC}}
\label{fig:Model 356A16}
\end{figure}

Another mode of measurement for natural frequency testing is a fiber optic accelerometer. These accelerometers are typically used in the power generation business as they are able to be placed inside generators while they are operational \cite{kalenik1998cantilever}. These are again typically used as monitoring devices, as their main purpose is to be used in a generator that is running. Being that they are fiber optic, this allows them to be close to a rotating field without the worry of noise or damage to the unit \cite{rong2017highly}. They are not typically used for a shaker or hammer test due to the fragile nature of the fiber optic cable. This type of measurement device senses the change in light being transmitted through the glass wire to the modulator. The modulator is then able to output the vibration level seen at the accelerometer position \cite{kalenik1998cantilever}. One example of a fiber optic accelerometer is a HBK model FS65ACC. This accelerometer is interesting in the fact that it specializes in low frequency applications from 0-50 Hz \cite{FOA}. This is a range that many accelerometers struggle with as it is very low in their operating range.

\begin{figure}[htbp]
\centering
\includegraphics[width=0.40 \linewidth]{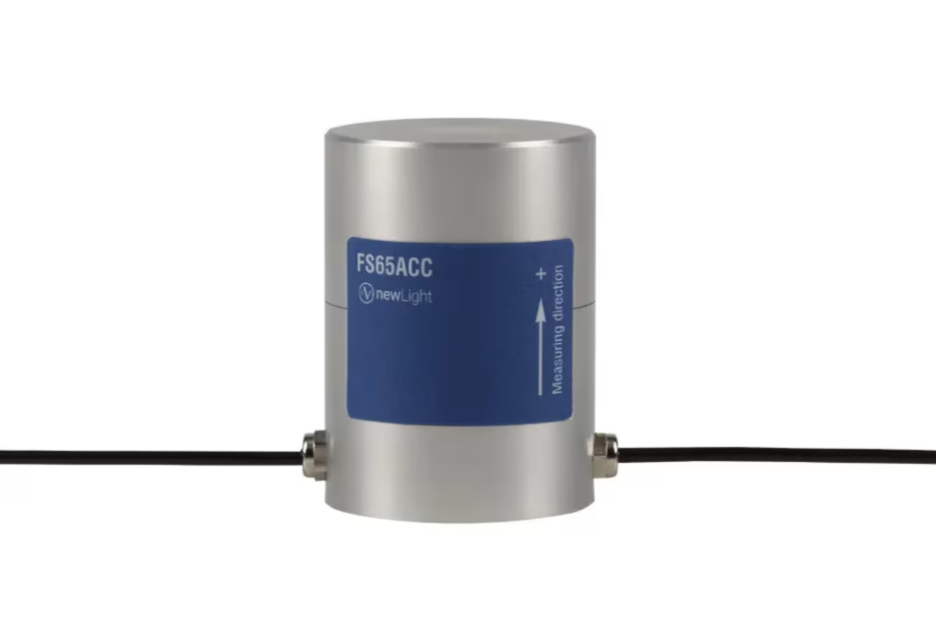}
\caption{Above is an example of a fiber optic accelerometer made by HBK Model FS65ACC \cite{FOA}}
\label{fig:Model FS65ACC}
\end{figure}

Velometer's are another option for reading vibration levels in natural frequency testing. A velometer, much like an accelerometer, measures vibration levels; the output data, however, are a separate measurement. An accelerometer measures the acceleration of an object and outputs the data in g’s/lb. Here, a velometer measures the velocity and outputs it as displacement/sec \cite{robert1996applications}. Although the velometer outputs a displacement/sec measurement, if we double-integrate this value, we are able to convert that value into g /lb. The way a velometer works is by having a small magnet inside the sensor that is fixed. Then, as the velometer starts to vibrate, a coil inside the velometer starts to produce an alternating voltage which can be correlated to the vibration levels seen at the device \cite{robert1996applications}. One example of a velometer is a model 330750 made by Baker Hughes \cite{Velomitor}. This specific velometer is used in high temperature applications, which can render itself useful in a piece of rotating machinery with a high operating temperature.    

\begin{figure}[htbp]
\centering
\includegraphics[width=0.50 \linewidth]{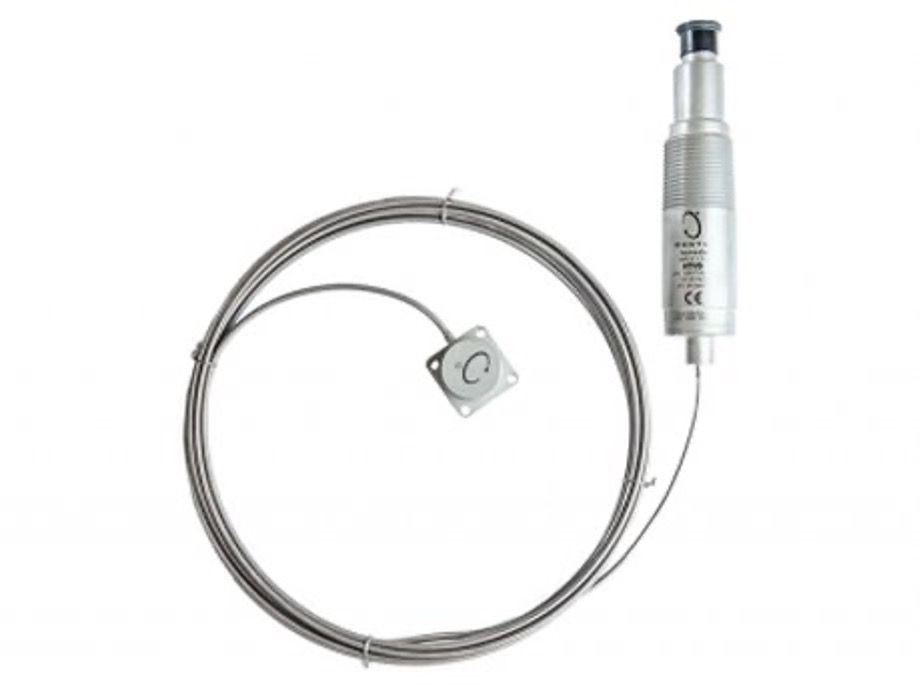}
\caption{Above is an example of a velometer, this example is made by Baker Hughes and is Model 330750. For this model the measurement head is the larger body which gets threaded into the measurement point of choice.     \cite{Velomitor}}
\label{fig:Model 330750}
\end{figure}

Mounting of all of these pieces of equipment is just as important as the type of instrument you decided to use. To mount a vibration device, a typical method is to use a strong magnet, which allows a rigid surface for the device to move with and will mimic the movement of the equipment \cite{VRMounting}. For our testing we used a PCB Piezotronics model 080A54 magnet which has a 50lbs/force rating \cite{Magnet}. In non-metallic objects, epoxies, glues, or beeswax are common forms of mounting techniques \cite{VRMounting}. These also offer a rigid surface that moves with the requirement to transmit the vibration levels to the device. Stud mounting is also a common practice used for more permanent applications where the measurement device will be placed in only a few selected areas \cite{VRMounting}. Placement of these devices should be placed in any areas of concern or areas that would help the viewer get a better understanding of the equipment being tested. This is important when looking at an ODS model, which is where you are able to see how the piece of equipment moves at certain resonant frequencies. Another thing to note as you choose your specific mounting form is the deviation that could be seen from your accelerometer. If we look at the Figure 1 we can see that our deviation shifts slightly depending on what mounting feature we opt to use \cite{PCBACC}. This specific deviation chart was pulled from the PCBs manual for model 356A16 \cite{PCBACC}. When choosing your measurement device and mounting structure it is best to consult the user manual for device specific recommendations.    

\begin{figure}[htbp]
\centering
\includegraphics[width=0.5 \linewidth]{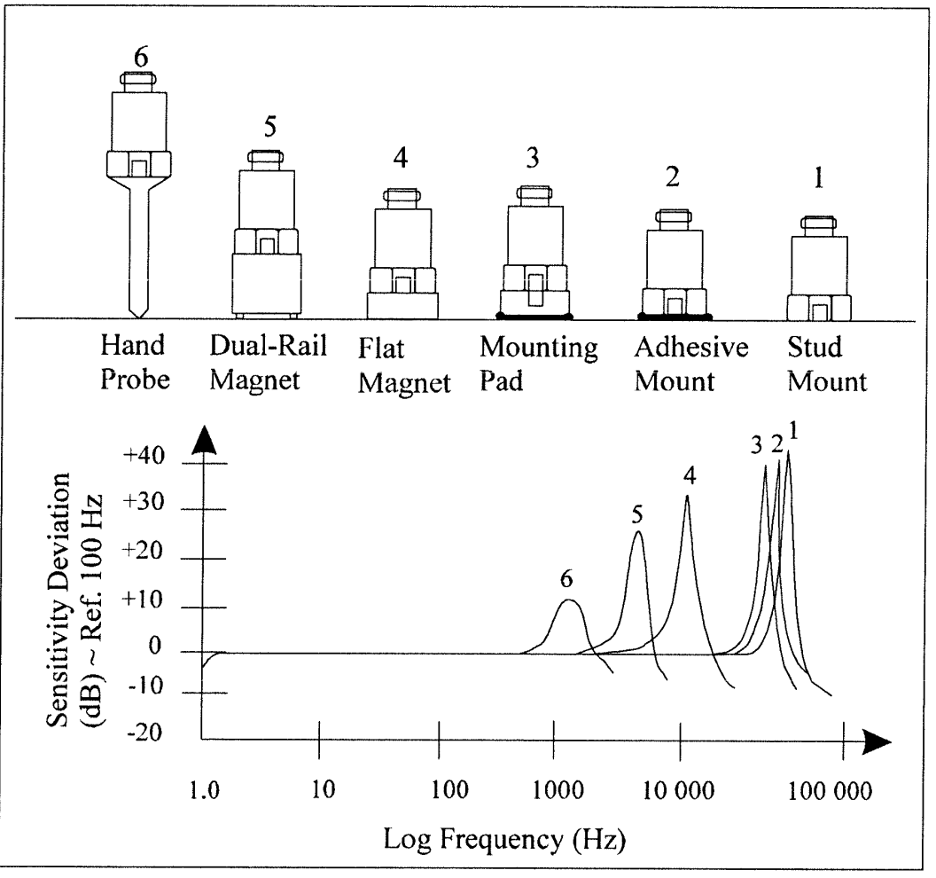}
\caption{This figure shows the sensitivity deviation of different accelerometer mounting options.  \cite{PCBACC}}
\label{fig:PCBACC}
\end{figure}

As mentioned for our gearbox testing, we opted to use a model 356A16 PCB tri-axial accelerometer, which was mounted via magnet \cite{PCBACC}. The model 356A16 has a ±10 percent sensitivity, a measurement range of ±50 g pk, and a frequency range of at least 0.5 to 4500 Hz on all three axis \cite{PCBACC}. Making this a more than suitable measurement device for our applications. As our frequency of interest of 0-200Hz and being a triaxial accelerometer allowed us to measure in three axes with one device. The utilization of a magnet made for quick point-to-point accelerometer moves and a firm structure to transfer accurate vibration levels to the device. Although our gearbox was metallic, our foundation during the test was not. In order to still utilize our magnets for the test, we employed an epoxy to mount steel washers to the foundation and grout plate. Once the epoxy was fully cured, it allowed us to conduct our test and measure the responses of the foundation without the need to employ two separate mounting methods. 

%%%%%%%%%%%%%%%%%%%%%%%%%%%%%%%%%%%%%%%%%%
\vspace{8pt} 

\section{Excitation Methods}

With your measurement device of choice in place, it is now time to excite the equipment in order to read the vibration response of the equipment. There are a few different methodologies used to excite a piece of equipment, which include: a shaker, impulse hammer, or running data when the equipment is in operation. In all of these equipment excitation methods, the vibration monitoring devices record the vibration levels seen at each measurement point. These points are triggered via detection of excitation in the case of the impulse hammer or through a trigger and then timed intervals, with the user defining the test parameters.   

Shaker units are a common form of excitation used in modal analysis. Shaker units come in a multitude of sizes and should be sized appropriately for the type of equipment that you are looking to test \cite{ashory1999high}. Shakers are used in applications when repeatability is important. A Shaker unit can output a certain amount of force at a specific frequency of interest. This makes this type of testing more repeatable as the excitation force and placement will be the same every time you test a piece of equipment \cite{ashory1999high}. An important characteristic of the shaker unit is that it excites the piece of equipment under test at one specific frequency. For this reason, a common practice is to do a swept frequency range. In this application, the shaker can be set up to start at a specific frequency, then slowly step up that frequency to the highest frequency of interest. This allows the operator to excite each frequency in that range to view any natural frequencies.

In Figure II, you will find a shaker which is being used for car door testing for an automobile. In this setup, the user is using an electric shaker unit to excite the door and then is using the three accelerometers to measure the frequency response. Shaker units can be electric or hydraulic, it is up to the user on which type of shaker they opt to use.

\begin{figure}[htbp]
\centering
\includegraphics[width=0.5 \linewidth]{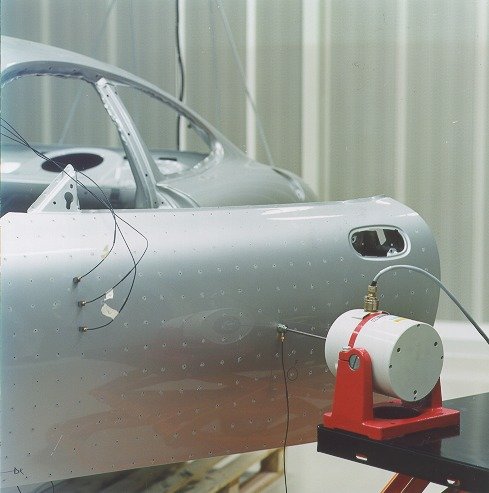}
\caption{This image depicts an electric shaker used in automotive vibration testing on a car door. \cite{ResearchGate}}
\label{ResearchGate fig}
\end{figure}

Things to take into consideration when looking to use a shaker unit are the time allotted for testing, accessibility of the equipment under test, and criticality of testing. Hydraulic shaker testing typically takes longer for setup and execution than a hammer test. With that in instances where there is limited time for testing, a hydraulic shaker test may not be the best option. Accessibility could also be an issue when looking to run a shaker test. Shakers tend to be much heavier and require the use of a hydraulic pump. With this requirement, the shaker test may not be the best-suited form of excitation. When implementing a shaker unit in your test, positioning of the unit can be critical to the accuracy of the data as well \cite{ashory1999high}. Natural frequencies are shifted utilizing weight; therefore, the implementation of a heavy unit may shift your resonant frequencies during the test, invalidating your test results \cite{ashory1999high}.  

Hammer excitation is another common form of excitation mode used in natural frequency testing. For this testing, a typical method is to use an instrumented hammer that has a quartz force sensor integrated with the head of the hammer. This force sensor is used to trigger the accelerometer once an impact is made to begin recording the responses \cite{avitabile2017modal}. Hammer natural frequency testing is unique from hydraulic shaker testing as the hammer allows for quicker data acquisition times. This quicker acquisition time is also coupled with a much more streamlined setup procedure makes this one of the more popular options for technicians and engineers performing this testing\cite{avitabile2017modal}. As this type of testing only requires a few cables and a hammer, eliminating the shaker and hydraulic pump. The drawback for this type of testing, compared to a shaker test, however, is the repeatability aspect of the testing. Hammer swings done by personnel can be ever so different, making predictably repeatable data difficult \cite{avitabile2017modal}. While these swings can also have defects in them, such as double impacts, overloading or under loading the excitation swing, or not hitting the equipment in the correct orientation \cite{avitabile2017modal}. 
	
Hammer selection in natural frequency testing is important to the data quality of the test at hand. Hammer size is important; this size is determined by the size of the equipment under test and the frequency range of interest \cite{avitabile2017modal}. Before selecting a hammer to use, it is always best to refer to the manufacturer’s hammer response curve to select a hammer size and tip type that best fits your application. A smaller hammer will result in a smaller nominal excitation level, whereas a larger hammer will have a larger nominal excitation level. For each hammer size, there are also different tips associated with the hammers. These tips are what determine your frequency range. A softer tip will excite a smaller frequency range \cite{ModelTLD086C02PCBpiezoelectronics}. Where a harder tip will excite a larger frequency range \cite{ModelTLD086C02PCBpiezoelectronics}. Frequency response curves put out by the equipment manufacturers are a good resource to utilize when deciding the best excitation hammer for their specific test. 

In instances where vibration levels of equipment are only of interest at running speed, a running data test may be performed. In this type of test, the excitation method is the equipment itself and does not require an outside source. When a piece of equipment is in an operating state or decoupled and in a simulated running state \cite{hermans1999modal}. This type of testing can be used to detect damaged or worn-out equipment. When testing under operating conditions, it can be easier to detect underlying problems that exist \cite{hermans1999modal}. This type of testing, much like the hammer and shaker testing, allows for ODS animations and FRF data to be obtained.  
	
For our testing we tested a custom made general electric gear box. For this test we opted to use a PCB TLD086D50 which is a 12lb impulse hammer with a medium to hard tip \cite{ModelTLD086C02PCBpiezoelectronics}. This hammer delivers a nominal impact of 1,000 lb/f, which is necessary for this large gearbox to ensure that we deliver a force large enough to be seen by our accelerometers \cite{ModelTLD086C02PCBpiezoelectronics}. Our medium to hard tip then allowed us to excite frequencies up from 0 to 900Hz \cite{ModelTLD086C02PCBpiezoelectronics}. In this specific test, we were interested in a range of 0 to 200Hz. This hammer selection made this response and measurement possible. Below you can find the response curve provided for the TLD086D50 by the manufacturer.

\begin{figure}[htbp]
\centering
\includegraphics[width=0.7\linewidth]{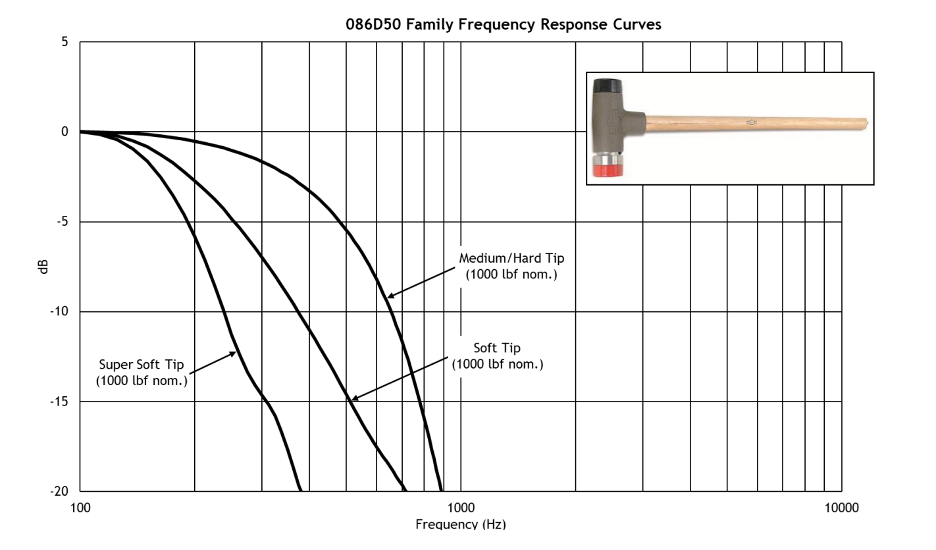}
\caption{Hammer response curve for model 086D50, which shows a caparison for frequency excitation capability for different hammer heads for this specific hammer. \cite{ModelTLD086C02PCBpiezoelectronics}}
\label{fig:ResearchGate}
\end{figure}

%%%%%%%%%%%%%%%%%%%%%%%%%%%%%%%%%%%%%%%%%%%%%%%%%%%   

\section{Data Gathered}

With the measurement devices placed and our form of excitation chosen and set up, we are now able to look at the types of data that we are able to process. In natural frequency testing, resonance frequencies are what most people look to identify with their test or analysis. This can be done in a number of different ways, such as looking at Frequency Response Function (FRF) graphs, using FRF's to look at the operational deflection shapes, or looking at the displacement seen on a given measurement device during test \cite{schwarz1999introduction}. In Figure III, you will find an example of an FRF curve with the resonant frequencies pointed to. Each spike on an FRF curve indicates a resonant frequency. While the phase indicates the motion in which the measurement device is moving. \cite{Siemens}

\begin{figure}[htbp]
\centering
\includegraphics[width=0.5\linewidth]{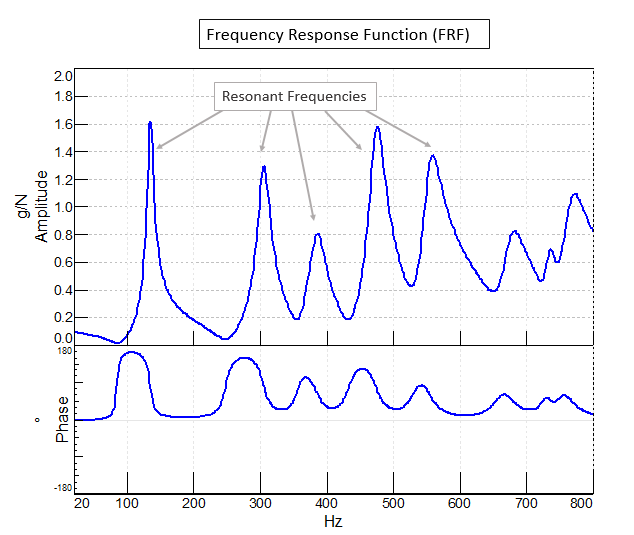}
\caption{FRF curve depicting several strong resonant frequencies being pointed to, along with the phase comparison to the measurement device responses.  \cite{Siemens}
}
\label{ResearchGate}
\end{figure}

Frequency Response Functions are graphed as g/N on the Y axis and frequency on the X axis. In this graph, the measurement device data is important, which will give you spikes in amplitude over your frequency range \cite{schwarz1999introduction}. These spikes indicate a resonance by the device on the piece of equipment. The higher the amplitude, the higher the response of that resonant frequency on the equipment \cite{schwarz1999introduction}. Although amplitude is a driving factor of the response significance. Another aspect that must be considered is the damping percentage of the response. In the world of natural frequency testing, damping can be used to identify the amount of time it will take to dissipate the response caused by the resonance \cite{lee2004evaluation}. When looking at an FRF curve, the wider the natural frequency response the larger the damping percentage will be \cite{Siemens}. When looking at equipment responses this is beneficial to identify as it will tell the operator how long it will take the equipment to dissipate this energy \cite{Siemens}. Conversely, it can tell you that your equipment will take a long time to dissipate this energy with a low damping percentage \cite{Siemens}. This can be important in rotating machines where your going to run through a natural frequency on your way to running speed. Then analysis and designers are able to look at the amplitude and the damping percentage to identify different design changes that need to be implemented to make the machine the most efficient.

Frequency Response Functions also lead to your operational deflection shape models. Frequency Response Functions can be viewed both in a gragh ORtable format or be translated into deflection models \cite{schwarz1999introduction}. In a deflection model, the modal analysis software takes your node responses from each measurement device, which translates that into the response seen on the equipment \cite{schwarz1999introduction}. This can be a quicker way to identify if different parts on the machine are out of phase \cite{schwarz1999introduction}. This could be correlated to a lost or damaged piece of the equipment. While also, it can tell the annalist which mode shapes they can expect out of the equipment \cite{schwarz1999introduction}.       

Displacement is another piece of information that we can receive from natural frequency testing. Using a PCB accelerometer, such as the one that we used for our gearbox vibration survey. We are able to take the g/N and double integrate this value to receive a displacement that is seen at the sensor \cite{schwarz1999introduction}. This can be used in running data tests, where the equipment is fully operational, and the operator is curious about what vibration levels are being seen at the machine. This data can also be looked at in comparison with frequency, which will give the user the ability to identify what frequency this displacement is occurring at. 

When the examiner examines this data, they are looking for a few different characteristics: mode shape, amplitude of response, damping percentage, measurement location, or node of the response \cite{schwarz1999introduction}. When annalists take these into consideration, they are then able to identify which modes of interest could be problematic for the equipment and how to mitigate them. When considering the natural frequency, if a natural frequency lands on a running speed or a known point at which vibration can be seen at that frequency. The owner is then able to shift this mode by means of adding weight to the equipment \cite{ccakar2018method}. This mass will then, in turn, shift the natural frequency down, whereas removing weight will shift this frequency up. Typically for equipment, it is easier and most efficient to add weight rather then take it off. However, the amount of weight you take off or add is equipment-specific and location-specific. Therefore, measurement locations and identification are important for analysis to know and understand. If a piece of equipment had a response of concern at the north end, adding mass on the south end would not shift the mode. While an analysis may also be interested in the damping percentage of the equipment. If the percentage is to low, they could try to add more damping features onto the equipment. This can be done by adding stiffening features such as elastomers which can help increase the damping of your equipment \cite{lee2004evaluation}. Or by adding mass springs or stiffener sprigs to the equipment \cite{ccakar2018method}. These tactics assist in dissipating the resonant frequency faster,which adds stability and longevity to the equipment \cite{ccakar2018method}.         

\section{Gear Box Testing Conditions}

In our testing, we measured the natural frequency of a gearbox that has been in service for 50 plus years. The reason for this testing was to determine whether the gearbox had experienced a shift in natural frequencies due to wear or if we could identify any loose parts that could be contributing towards higher-than-normal vibration levels during operation. In doing so, we also experimented with test conditions in order to better understand the effects of different conditions during the test. To do this we used the Siemens LMS software package and SCADAS module. This allowed us to create an ODS modal to view the responses of our accelerometers and to watch what responses we had at the identified natural frequencies. Below you will find our ODS modal we used for this test. As well as tables which dictate which part of the gearbox each point is in. Points 33, 34, and 35 were supporting piping that were not excited by our impacts on the gearbox.        

\begin{figure}[htbp]
\centering
\includegraphics[width=1 \linewidth]{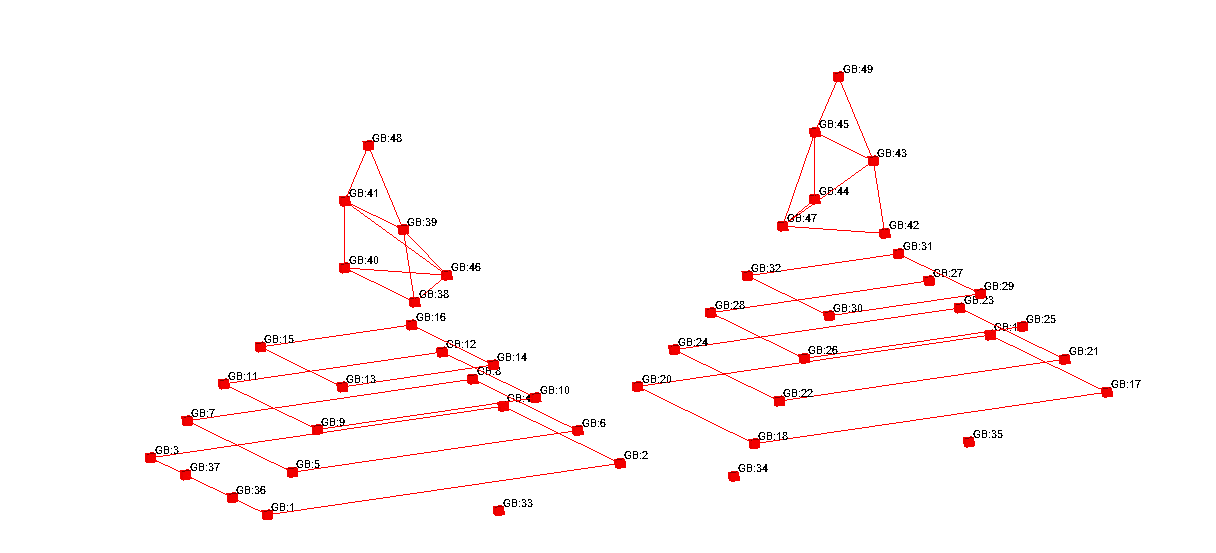}
\caption{ODS Model for the gear box under test, where the left side, when in operation, is being driven by the motor. While the right side is what drives the object you wish to rotate.}
\label{fig:ODS Modal}
\end{figure}

\begin{table}[ht]
    \centering
    \footnotesize
    \caption{Point descriptions for gear box pedestal 1 points, left pedestal from figure 8}
    \label{Gear Box Pedestal 1}
    \begin{tabular}{|c|c|c|c|}
        \hline
        \multicolumn{4}{|c|}{Gear Box Pedestal 1} \\
        \hline
        Gear Box Foundation & Grout Plate & Steel Plate & Gear Box \\
        \hline
        1  & 5 & 9  & 13 \\\hline
        2  & 6 & 10 & 14 \\\hline
        3  & 7 & 11 & 15 \\\hline
        4  & 8 & 12 & 16 \\\hline
        36 &   &    & 38 \\\hline
        37 &   &    & 39 \\\hline
           &   &    & 40 \\\hline
           &   &    & 41 \\\hline
           &   &    & 46 \\\hline
           &   &    & 48 \\
        \hline
    \end{tabular}
\end{table}

\begin{table}[ht]
    \centering
    \footnotesize
    \caption{Point descriptions for gear box pedestal 2 points, right pedestal from figure 8}
    \label{tab:placeholder_label}
    \begin{tabular}{|c|c|c|c|}
        \hline
        \multicolumn{4}{|c|}{\textbf{Gear Box Pedestal 2}} \\
        \hline
        Gear Box Foundation & Grout Plate & Steel Plate & Gear Box \\
        \hline
        17 & 21 & 25 & 29 \\\hline
        18 & 22 & 26 & 30 \\\hline
        19 & 23 & 27 & 31 \\\hline
        20 & 24 & 28 & 32 \\\hline
           &    &    & 42 \\\hline
           &    &    & 43 \\\hline
           &    &    & 44 \\\hline
           &    &    & 45 \\\hline
           &    &    & 47 \\\hline
           &    &    & 49 \\
        \hline
    
    \end{tabular}
\end{table}

\section{Noise Levels}

Unless in a controlled lab environment, noise levels can be difficult to overcome. Therefore, for our test, we took the opportunity to understand what effects a lube oil system would have on the test that we were conducting. To do this, we completed the test twice, once with the lube oil system on and then again with the system off. This now gives us a clear comparison on the effects of the lube oil system and what effects this type of noise has on our data. 

In figures 9 and 10 you will find two graphs, one with the lube oil system on and the other with the lube oil system off. As we can see, there are sharp spikes being read from our accelerometers at two septate frequencies. This is a direct effect of our lube oil system. The other thing to note is the identification of stable curves in the two stabilization curves \cite{Stabilization}. For this test, we utilized the Siemens LMS system; in this software package, we are able to sum all data taken from our accelerometers. From there, we are able to create a spectral plot to identify peaks which represent natural frequencies \cite{Stabilization}. The data obtained for figure 9 and 10 was from this summed stabilization curve. As we can see in Figure 9, we have three strong spikes that are not seen when testing with the lube oil system off. Which drowns out some of the other peaks that we would have seen if the system was of. As we are only able to see the spikes from the operating speeds of the motors.  Therefore, we can conclude from this test that, when possible, all supporting equipment should be turned off when conducting a natural frequency test. Failure to do so has the potential to result in spikes in the data or cause the software to not be able to identify modes.   

\begin{figure}[htbp]
\centering
\includegraphics[width=.9 \linewidth]{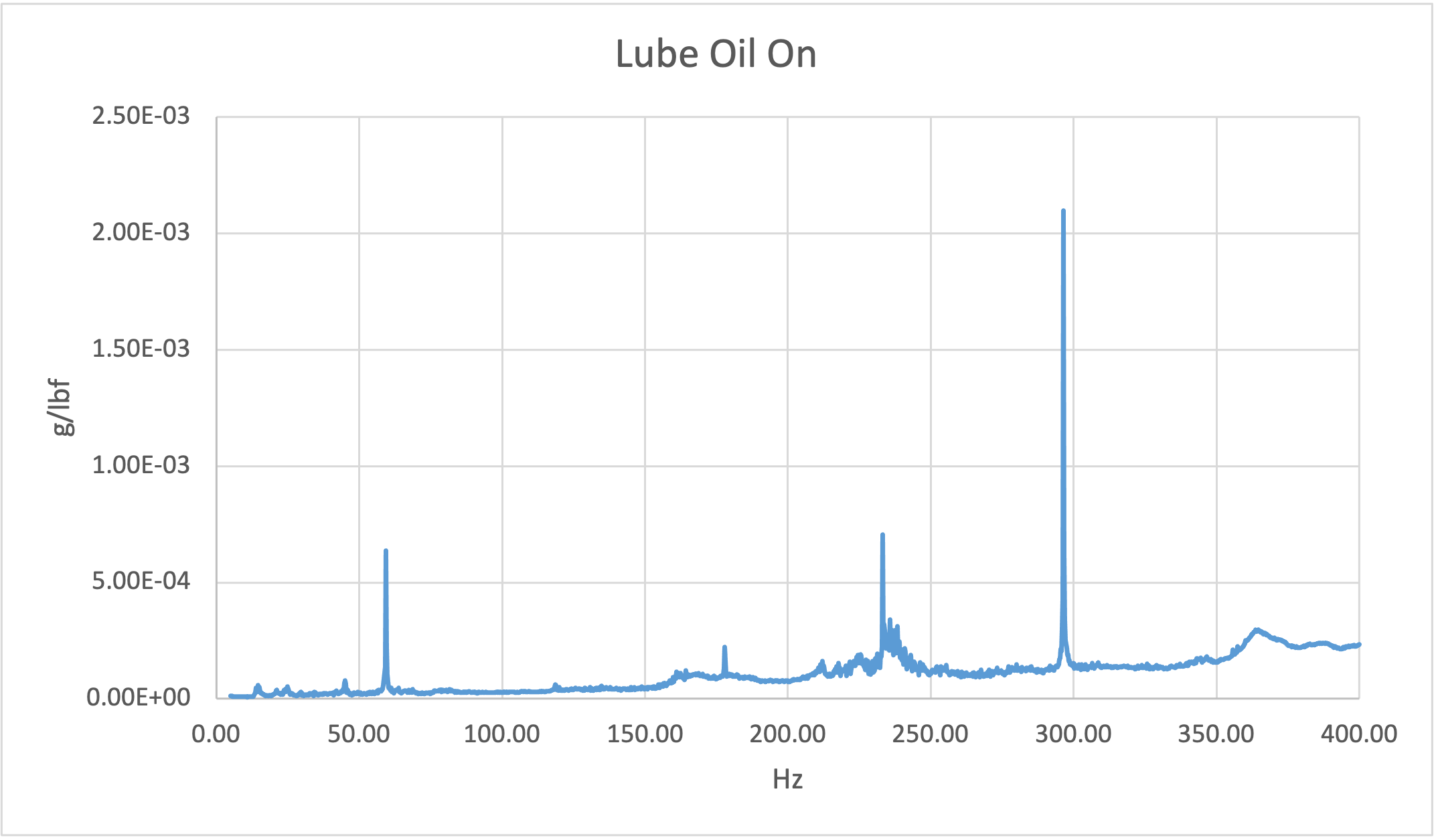}
\caption{Spectral Plot when lube oil is on, which shows three strong peaks for operating speeds of the pumps}
\label{fig:Lube Oil on}
\end{figure}

\begin{figure}[htbp]
\centering
\includegraphics[width=.9 \linewidth]{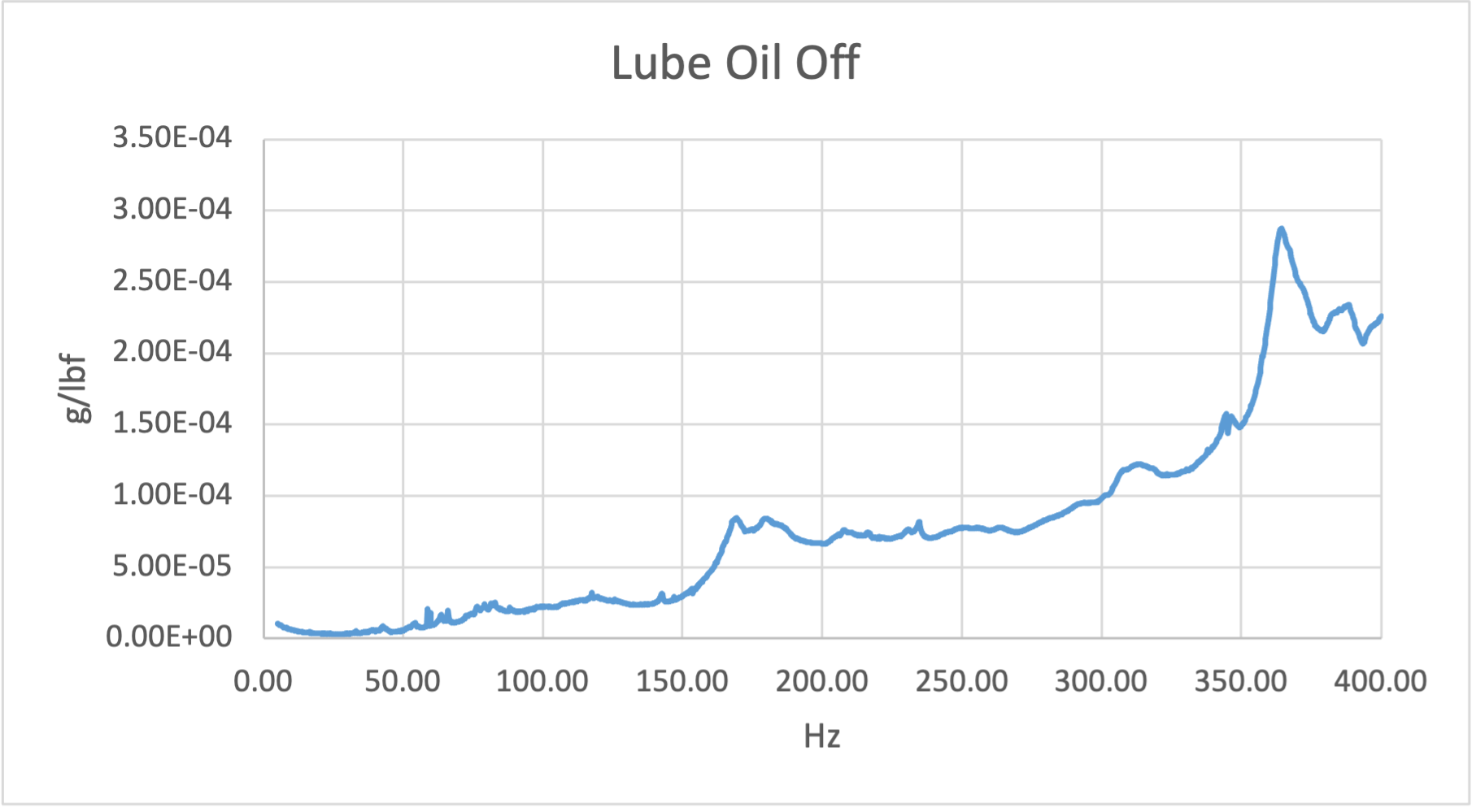}
\caption{Spectral Plot with Lube Oil off}
\label{fig:Lube Oil off}
\end{figure}

\section{Excitation Points}

When conducting a hammer test with a triaxial accelerometer, it is typical to excite in all three directions. This can be done by impacting on an angle, which then excites all three directions. Alternatively, you are also able to excite these axes individually to ensure that each axis gets properly excited. For our gearbox testing, we opted to excite the gearbox in all three axes. To ensure that we get proper excitation in all three axis. On top of this, we also wanted to experiment with different points at which we decided to impact with our hammer. To do this, we had an excitation point directly on the gearbox, while the other was on the foundation of the gearbox. In doing so, we are able to compare the data results from the two hammer locations to examine if hammer location does, in fact, play a part in the effects of excitation.

In the graph depicted below, we can compare two different excitation points as well as the summed value of all excitation data. The orange curve is from the impact on the gearbox itself. Which we can see is the largest response of all the curves. The gray curve, where we impacted the foundation, has the least response out of all the excitation data. From there, the blue curve that has summed data from both impact locations sits just about in the middle of the two curves. All curves share a similar shape and respond only with differing amplitudes. Which tells us that even when we impact in less ideal locations during a test, we should still get a response; it may just be damped. In many types of natural frequency testing software, you are able to range the measurement device. After examining this data and looking at the results, it could be advantageous to use this feature if switching between an optimal and less optimal excitation point. Or due to field conditions you are restricted to only be in less advantages impact locations.

\begin{figure}[htbp]
\centering
\includegraphics[width=.9 \linewidth]{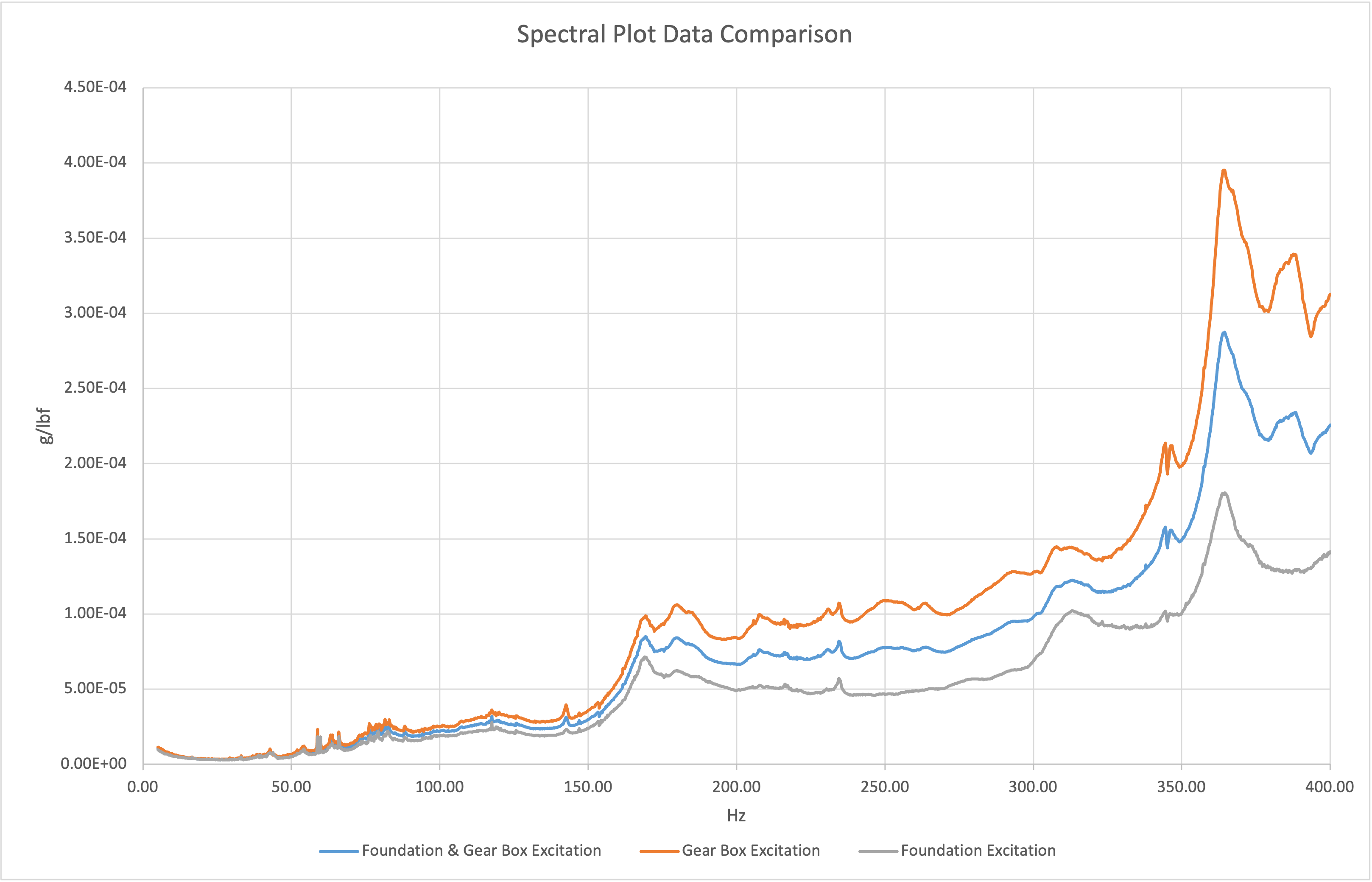}
\caption{Spectral Plot with Foundation Excitation Point in Grey, Gear Box Excitation in Orange, and Both Foundation and Gear Box Excitation in Blue}
\label{fig:Swpectral plot comparison}
\end{figure}

Now if we look at the graph below we can again look at the data taken with the lube oil system off and both excitation points. As we can see the stabilization curve response from the combined excitation points lands somewhere in the middle of the two. This makes sense, as this graph is a summed plot of all available data being captured. This brings the question of what would have been the best excitation point used for this test, and whether it is better to use both the foundation and gearbox responses. After examining the data, it is less critical to use both the foundation and gearbox housing excitation points. As one is able to excite all three axes of the accelerometers and give an accurate response. Although both points resulted in similar responses, it would be more advantageous to impact the form of excitation during operation. Meaning that during operation, the gearbox will have excitation internally, therefore the housing is more applicable than the foundation as a form of excitation.

\section{Testing on a Welded Seam}
The last smaller test that we conducted was the effects of testing near a welded seam on the equipment. The goal of this test was to validate the placement of an accelerometer on and near a weld on a piece of equipment. By utilizing an ODS modal, we are able to prove that the two accelerometers placed on either side of a welded seam moved the same distance both in displacement and in phase. This proves that when testing on a seam, the side of the weld does not matter and that less care needs to be taken when deciding instrumentation placement. However, prior to placing instrumentation and conducting testing, inspection of welds is advised. In the case where there should be a poor weld or a damaged weld, it could be advantageous to still test on both sides of a seam to ensure accurate results.

\section{Conclusion}
In the conclusion of the gearbox testing, it was found that there were no natural frequency concerns that would have resulted in the high vibration. However, this testing resulted in a great opportunity to look at the effects of testing on a welded seam, excitation point choosing, and the effects of external noise. This real-world testing gives us the insight and ability to plan future tests out more efficiently and confidently.

%%%%%%%%%%%%%%%%%%%%%%%%%%%%%%%%%%%%%%%%%%
%\isPreprints{}{% This command is only used for ``preprints''.
%\begin{adjustwidth}{-\extralength}{0cm}
%} % If the paper is ``preprints'', please uncomment this parenthesis.
%\printendnotes[custom] % Un-comment to print a list of endnotes

%\reftitle{R}
    \bibliographystyle{ieeetr}
    \bibliography{references}

%\PublishersNote{}
%\isPreprints{}{% This command is only used for ``preprints''.
%\end{adjustwidth}
%} % If the paper is ``preprints'', please uncomment this parenthesis.
\end{document}